\documentclass[letterpaper, 10 pt, conference]{ieeeconf}  

\IEEEoverridecommandlockouts                              
\usepackage{url}
\pdfoutput=1
\usepackage{graphics} 
\usepackage[T1]{fontenc}
\usepackage{color}
\usepackage{graphicx}
\usepackage{float}
\usepackage{verbatim}
\usepackage{cite}
\title{\LARGE \bf
A Deep Neural Network to identify foreshocks in real time 
}
\author{K.Vikraman}%

\author{K.Vikraman  \\{\tt\small mail@vikramank.com } \\{\tt\small vikramaniitr@gmail.com }
}

\begin{document}

\maketitle
\thispagestyle{empty}
\pagestyle{empty}

\begin{abstract}
Foreshock events provide valuable insight to predict imminent major earthquakes. However,
it is difficult to identify them in real time. In this paper, I propose an algorithm based on deep learning to instantaneously classify a seismic waveform as a foreshock, mainshock or an aftershock event achieving a high accuracy of 99\% in classification. As a result, this is by far the most reliable method to predict major earthquakes that are preceded by foreshocks. In addition, I discuss methods to create an earthquake dataset that is compatible with deep networks. 
\end{abstract}

\section{INTRODUCTION}

Major earthquakes cause massive destruction to life and property. To mitigate the devastation from these events we need to find a reliable earthquake prediction method. There have been several research studies in the past to predict a major earthquake. However, there is no reliable method to predict major earthquake that is imminent \cite{geller1997earthquake}.

 Nevertheless, there are few successful predictions. The Haicheng earthquake(1975) is a perfect example for successful prediction. There were different types of precursors\cite{wang2006predicting} for this event including a foreshock sequence, peculiarity in animal behavior and anomalies like geodetic deformation, differences in groundwater level. But these abnormalities were not present in other major earthquakes. Hence there is no universal precursor for earthquakes. 
 
 However, a majority of the earthquakes(M> 7.0Mw) are preceded by foreshock sequences, making them a good precursor. Foreshocks are earthquakes that precede major earthquakes. Foreshocks can be a warning signal that a major earthquake is about to happen in that vicinity. For instance, in March 2011, a powerful foreshock of magnitude 7.3Mw struck the North East coast of Japan that was followed by a major earthquake of 9.0Mw two days later. Currently, foreshocks are indiscernible from other earthquake types when the event occurs\cite{ogata2014comparing}.
 
In this paper, I discuss a powerful algorithm that can identify and classify earthquake events into foreshock, mainshock or aftershock event instantaneously based on seismic waveform data. I am using deep learning approach to train and teach my algorithm about different types of earthquakes particularly foreshock events. 

The paper is organized into five sections. In the first section, I discuss types of earthquakes and difficulties in distinguishing them in real time. In second section, I introduce deep learning algorithm followed by methods to create earthquake dataset. In the final two sections I discuss about my network architecture and analysis of the performance of my algorithm.
\section{Types of earthquakes}
\subsection{Foreshocks}
Forshocks are earthquakes that precede major earthquakes and appear in the vicinity of them. The magnitudes of earthquakes ranging from as low as 3Mw upto as high as 7.6Mw. They occur in a random fashion; some of them appear just few hours before the main event; but very few occur months and infact years before the main event.  Majority of foreshocks are reported to arrive within 24 hours of the main event.
\\
Challenges in identifying foreshocks: 
\begin{itemize}
\item Sometimes indistinguishable from other events\cite{ogata1996statistical}
\item Number of foreshock events in a sequence varies\cite{chen2013california}
\item Classified into different types \cite{helmstetter2002foreshocks}
\item Occur in varied time interval.Some might occur much ahead of the main shock. For example, the foreshock of Sumatra earthquake(2004) is said to have occurred two years earlier with a magnitude of 7.3Mw.
\item Occur in varied magnitudes. For instance, 2011 earthquake in Miyagi Japan witnessed a powerful foreshock of magnitude of 7.3Mw
\item Cannot be identified as a foreshock in real time.\cite{ogata2014comparing}
\end{itemize}

\subsection{Mainshocks}
Mainshocks are events that are either individual or cluster of events. Minor earthquakes with magnitude less than 4.0Mw are generally individual events. On the contrast, major earthquakes of magnitude 6.0 and above are generally accompanied by a series of aftershocks. Also majority of the main shocks are preceded by foreshocks. 

\subsection{Aftershocks}
Aftershocks are earthquakes that follow a major earthquake. They release the left out energy and occur as a sequence of events. Unlike a foreshock and a mainshock event, they always occur as sequence of events and can last up to a few days. Though aftershock events cluster around the mainshock event, their impact can be felt  at greater distances\cite{parsons2014global}.

\section{Deep Learning}

Deep learning is the state of the art Machine learning algorithm. It has performed well in terms of classification tasks of Images, speech signals, etc. Deep learning algorithms translate the indistinguishable input data into higher dimensional space where they will be linearly separable \cite{colah}. \\
Typical deep networks consist of few layers of convolutional neural network(CNN) followed by couple of fully connected layers. Each of CNN has several filters/kernels that extract features of the network. In addition they consist of pooling layers to pool similar features. 
\subsection{Important advantages of Deep networks\cite{lecun2015deep}}
\begin{itemize}
\item \textbf{Representation Learning}\\
The network can understand representations of data with multiple levels of abstraction

\item \textbf{Automatic feature extraction} \\
The network can extract essential features of a raw data without manually hand engineering them and doesn't require knowledge expertise in the relevant field  

\item \textbf{Invariance} \\
The network is invariant to small variations in occurrence in time or position 
\end{itemize}

 \subsection{Deep learning and seismology}
 This paper is one of the first applications of deep learning in the field of seismology and earthquake prediction.
In real time, it is difficult to identify and classify events. Given the limited time to prove the usefulness of foreshock events and to overcome all the challenges discussed in the previous section, I created an algorithm based on deep learning 
that takes the seismic data as input and outputs the possible event type.

\section{Dataset}

To harness the power of deep learning; it requires huge labeled dataset like MNIST, CIFAR10. At present there are no datasets available that are required for this study. Hence I created my own dataset. 

\subsection{Data acquisition}
\begin{figure}[ht]
\centering{\includegraphics[scale=0.4]{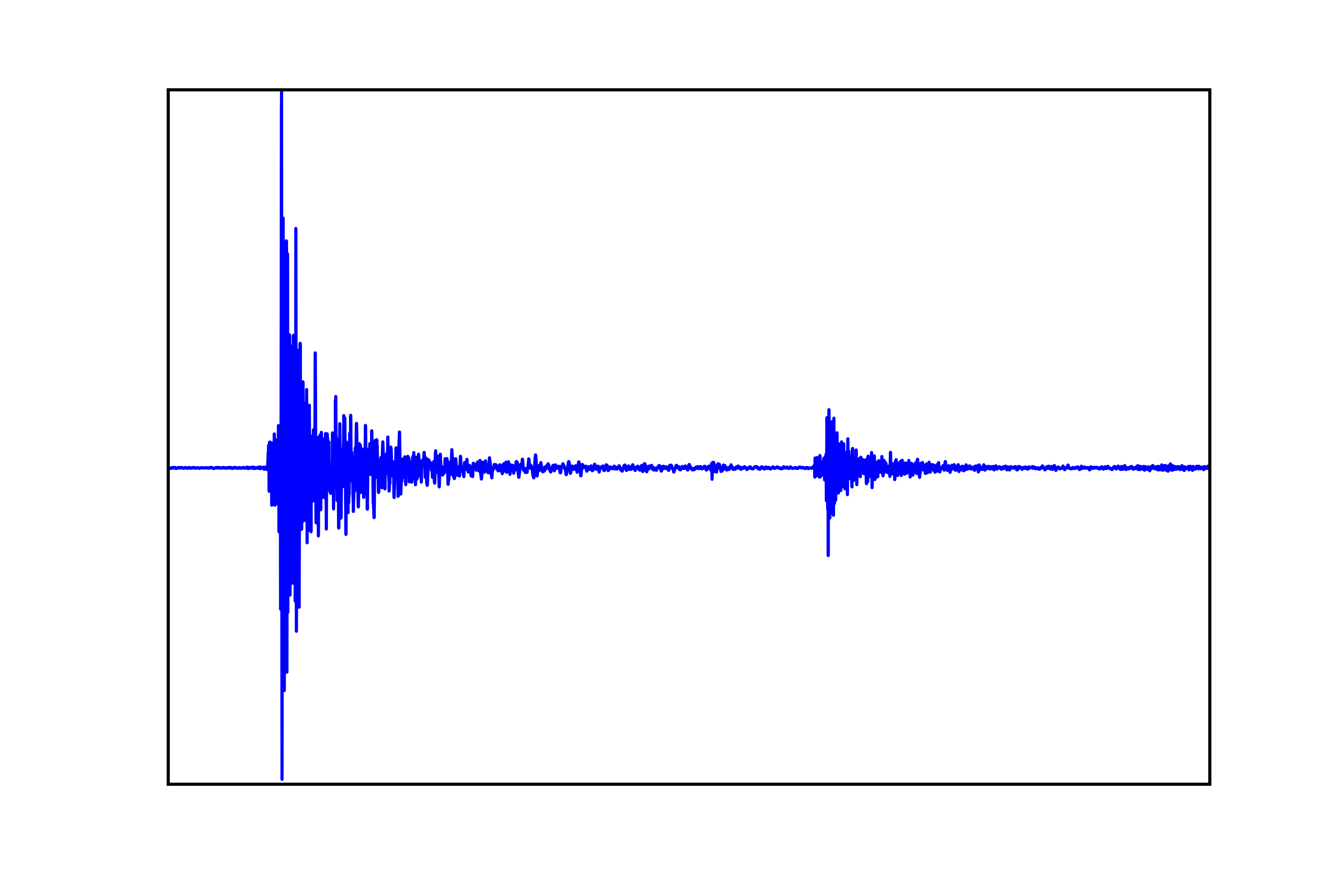}}
\caption{Sample Waveform of a seismic data in time domain}
\centering
\end{figure}
Seismic waveforms are readily available. In my analysis I used seismic waveforms as an input data. 
Throughout my analysis I used time-series waverform data of earthquakes. I extracted the data from wilber3, IRIS \cite{iris}. To maintain uniformity, I used data from station that is closest to that particular event and data points from one minute prior to P-wave arrival till 10 minutes after P-wave arrival. I also restricted to collecting data from 'BH' channel alone.
\subsubsection{Foreshock data}
Foreshock events can be identified using algorithm described in \cite{ogata1996statistical} or some papers \cite{chen2013california} that explicitly mention event details. I gathered data from the following 
events and in total I collected 191 foreshock events.
\\

\begin{table}[ht]
\centering{\caption{Foreshock data}
\begin{tabular}{|c | c | c |}
\hline
S.no& Foreshock event(s) & Magnitude \\ 
& & (of corresponding mainshock) \\
\hline
1 &  El Mayor-Cucapah \cite{chen2013california} & 7.2Mw \\ 
& CA, USA & (4/4/2010) \\
\hline
2 &  Hector Mine \cite{chen2013california} & 7.1 Mw  \\ 
& CA, USA & (16/10/1999) \\
\hline
3 &  Landers \cite{chen2013california}  &  7.3 Mw \\ 
& CA, USA & 28/6/1992 \\
\hline
4 & Kumamoto   &  7.0 Mw  \\ 
& Japan & (16/04/2016) \\
\hline
5 &  Miyagi & 7.0 MW  \\ 
&Japan & (11/03/2011) \\
\hline

\end{tabular}}
\end{table}

\subsubsection{Mainshock data} 
In my dataset I collected mainshocks of magnitude above 7.2Mw. Further, to introduce variety I included 25 individual mainshock events from seismically active places like Okinawa.  Overall I collected 224 mainshock events. 
\subsubsection{Aftershock data}
I took aftershock events that followed major earthquakes in Nepal(2015) and New Zealand earthquake(2016).Further, I collected aftershock data from the which I gathered foreshock data, viz.  Miyagi (2011) and Kumamoto (2016), Japan. In total, I have gathered 224 aftershock events. 
\subsection{Data preprocessing}
IRIS Wilber 3 provides bundled data and provides in 'SAC' format. The seismic data is in time domain. To convert this into useful information I transformed the input data to frequency domain. I used python package ObsPy\cite{wassermann2013obspy} to extract the stored data and SciPy\cite{jones20012011} to perform Fast Fourier Transform (FFT) of the signals. 

\begin{figure}[ht]
\centering{\includegraphics[scale=0.45]{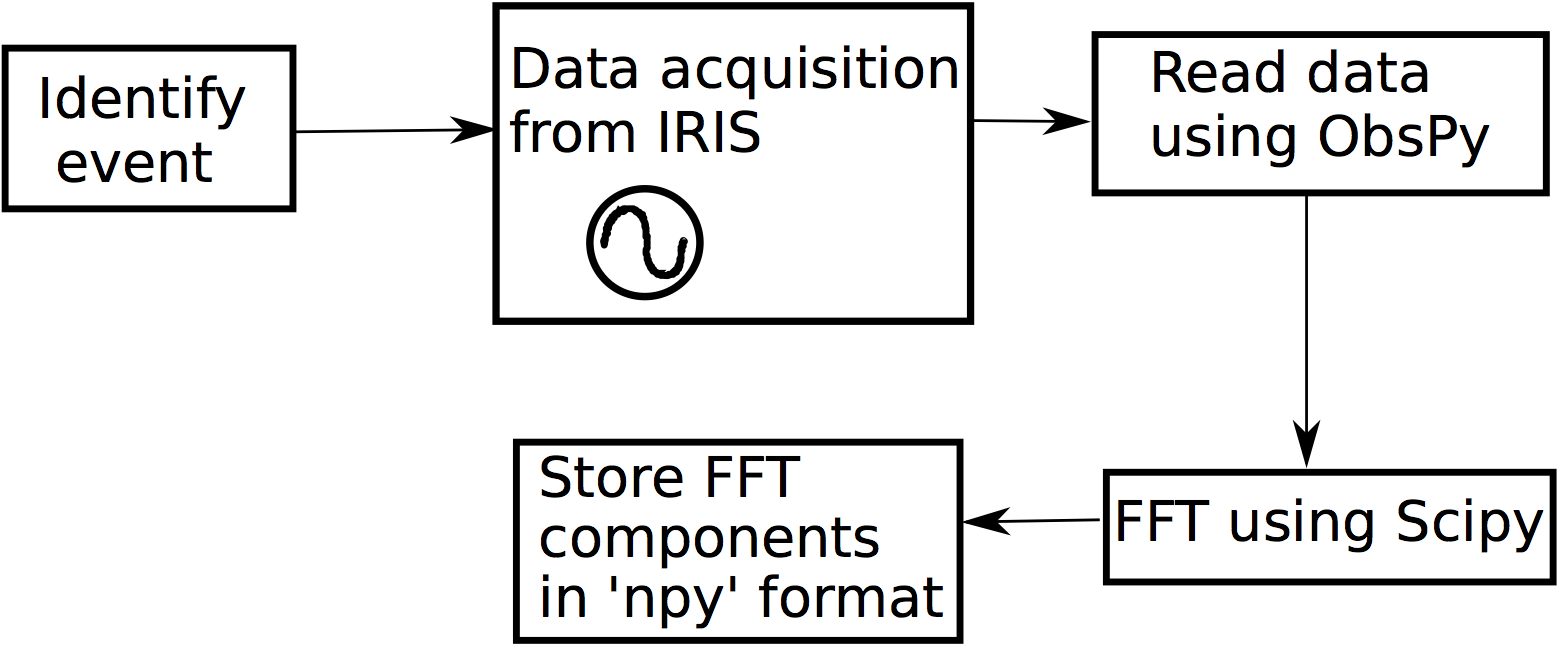}}
\caption{Data acquisition and processing}
\centering
\end{figure}
\section{Network architecture}
 I have created a deep feed forward network. Since my dataset is labeled, this is a supervised learning algorithm. I used a variant of AlexNet\cite{krizhevsky2012imagenet} with the following configuration
\begin{itemize}
\item Four convolutional layers
\item Two fully connected layers
\item Xavier weight initialization
\item Non-linear activation function ReLU 
\item ADAM Optimiser 
\end{itemize}
\begin{figure}[ht]
\centering{\includegraphics[scale=0.45]{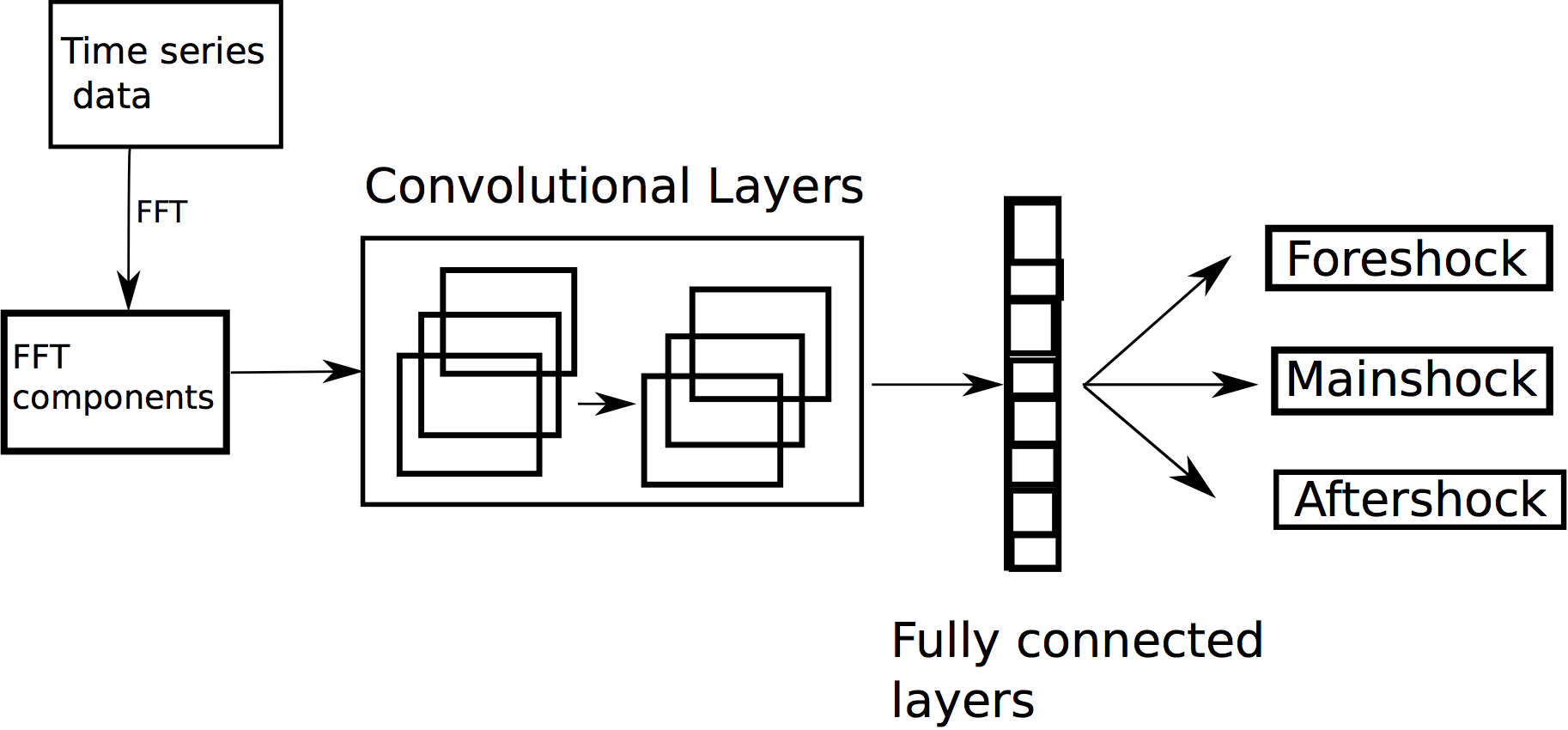}}
\caption{Network architecture}
\centering
\end{figure}
The network takes FFT components of time-series data of the seismic waveforms  as inputs and outputs the class in which individual Fourier component belongs to. There are three output classes corresponding to foreshock, mainshock and aftershock respectively.
In order to classify an entire event, I take the take the output of the network and count the number of outputs per class. The class with the highest number of counts corresponds to the predicted event.  
\subsection{Software}
In this study, I used a popular deep learning framework TensorFlow. It has a native interface for Python and is robust while deployed in larger scale. Since, the main purpose of this project is for real time deployment, TensorFlow is the best choice of framework.
\subsection{Hardware}
Generally deep learning algorithms require huge computational power including high RAM and GPUs. In my case, though the computations are huge, my data is one-dimensional and hence takes relatively lesser time and computation. I used Quad Core CPU with 8GB RAM for my computations.

\section{Analysis}
The hyper-parameters I chose are number of filters in CNN layer and number of neurons in the fully connected layer. I incorporated dropout layers\cite{srivastava2014dropout} and tried 'L1/L2' loss as a regulariser to boost validation set accuracy.
\begin{figure}[ht]
\centering{\includegraphics[scale=3.5]{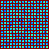}}
\caption{ Visualization of select filters from first layer}
\end{figure}

\subsection{Accuracy}
I first report accuracy in predicting individual Fourier components of particular waveform. After training the network for 120 iterations overall accuracy in the training dataset is 98.6\% and validation dataset is 92.4\%. Augmenting the dataset would bridge the gap between training accuracy and test accuracy.
\begin{figure}[ht]
\centering{\includegraphics[scale=0.5]{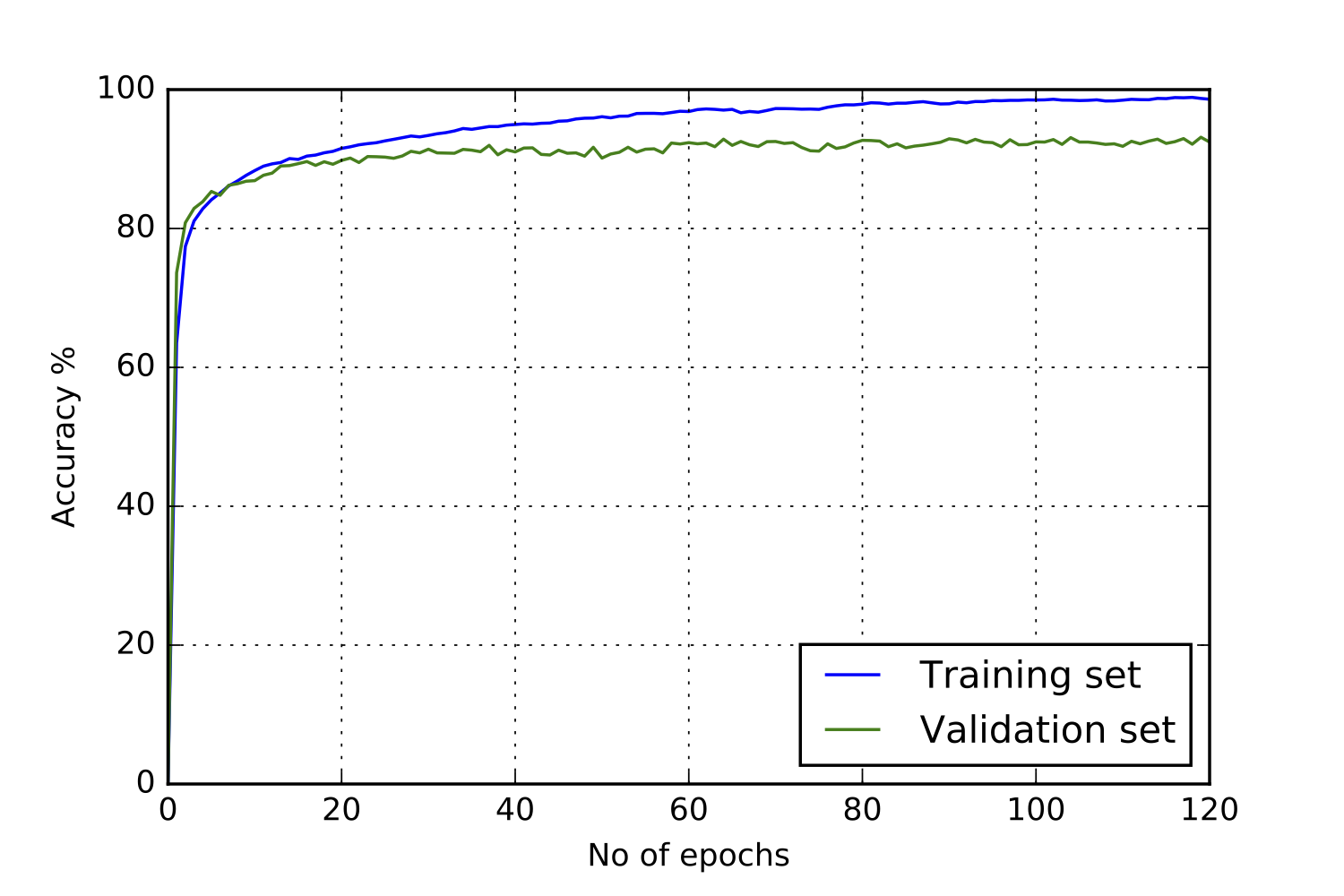}}
\caption{Overall accuracy of predicting the components from FFT}
\end{figure}
\\
\subsection{Confusion Matrix}
Confusion matrix provides valuable information about the incorrect classification and further analysis. From the figure, we can notice that foreshock and aftershock share some common properties.
\begin{figure}[ht]
\centering{\includegraphics[scale=0.4]{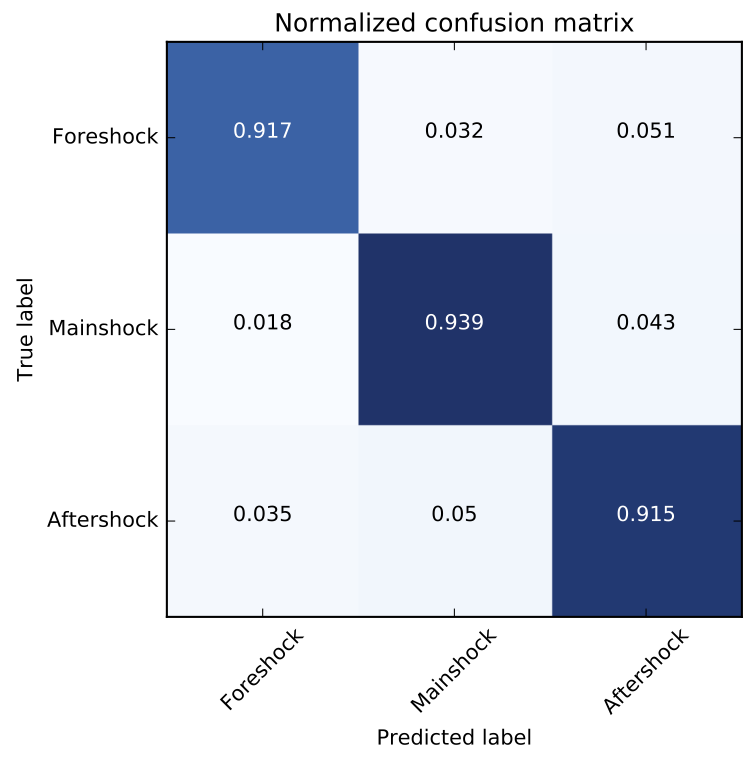}}
\caption{Confusion matrix of the validation set}
\centering
\end{figure}
\\
\subsection{Accuracy on individual events}
\textbf{Hypothesis}\\
Every event has several components. The network returns the possible class that a particular component belongs to. As mentioned earlier, the final predicted class it the class with maximum number of components in it. For example, if I take the time series data of a event 'X', the FFT will return 'y' components. Now feeding these y components to the network will return y outputs with respective predicted class. Let's say 'a','b' and 'c' number of outputs correspond to foreshock, mainshock and aftershock respectively. The final prediction will be the class that has maximum number of outputs, i.e, if 'a' is the greatest among the three, the predicted event will be foreshock event. Based on this definition, I have calculated the accuracy of classification of all the events. \\
The deep network is able to classify foreshock and aftershock events with 100\% accuracy.

\begin{table}[h!]
\centering{\caption{Accuracy on overall event classification}
\begin{tabular}{|c | c | c |}
\hline
S.no& Event & Accuracy  \\ 
\hline
1 & Foreshock&  100\% (191/191)\\ 
\hline
2 & Mainshock & 99.5\% (223/224) \\ 
\hline
3 & Aftershock & 100\% (224/224)  \\
\hline
\end{tabular}}
\end{table}

\section{CONCLUSIONS}
In this paper, I have demonstrated that foreshocks can be identified in real time with great accuracy. To achieve the true success of this algorithm, one needs to augment dataset I created, by including a variety of earthquake events.

\section{Future work}
The results from my work are really promising. I would like to work along with USGS and IRIS to create more earthquake datasets and apply deep learning algorithms to understand more about the behavior of earthquakes.
 
\addtolength{\textheight}{-12cm}   



\section*{ACKNOWLEDGMENT}
I would like to thank Parag K Mittal from CADL, Kadenze \cite{parag} for creating great video lectures and 
tutorials about Deep learning, my friend Arushi Saxena(University of Memphis) for providing feedback and
valuable information about seismology, Adam Clark from IRIS \cite{iris} for his assistance in creating the data  
and my family for their love and support.


\bibliography{final.bbl}

\begin{thebibliography}{10}

\bibitem{iris}
Iris: Wilber 3: Select event.
\newblock \url{http://ds.iris.edu/wilber3/find_event}.
\newblock (Accessed on 11/26/2016).

\bibitem{colah}
Neural networks, manifolds, and topology -- colah's blog.
\newblock \url{http://colah.github.io/posts/2014-03-NN-Manifolds-Topology/}.
\newblock (Accessed on 11/26/2016).

\bibitem{parag}
Online arts and technology courses | kadenze.
\newblock \url{https://www.kadenze.com/}.
\newblock (Accessed on 11/26/2016).

\bibitem{chen2013california}
Xiaowei Chen and Peter~M Shearer.
\newblock California foreshock sequences suggest aseismic triggering process.
\newblock {\em Geophysical Research Letters}, 40(11):2602--2607, 2013.

\bibitem{geller1997earthquake}
Robert~J Geller.
\newblock Earthquake prediction: a critical review.
\newblock {\em Geophysical Journal International}, 131(3):425--450, 1997.

\bibitem{helmstetter2002foreshocks}
A~Helmstetter and D~Sornette.
\newblock Foreshocks and earthquake predictability.
\newblock {\em JB}, 2409(physics/0210130):01, 2002.

\bibitem{jones20012011}
E~Jones, T~Oliphant, P~Peterson, et~al.
\newblock 2011. scipy: opensource scientific tools for python, 2001.

\bibitem{krizhevsky2012imagenet}
Alex Krizhevsky, Ilya Sutskever, and Geoffrey~E Hinton.
\newblock Imagenet classification with deep convolutional neural networks.
\newblock In {\em Advances in neural information processing systems}, pages
  1097--1105, 2012.

\bibitem{lecun2015deep}
Yann LeCun, Yoshua Bengio, and Geoffrey Hinton.
\newblock Deep learning.
\newblock {\em Nature}, 521(7553):436--444, 2015.

\bibitem{ogata2014comparing}
Yosihiko Ogata and Koichi Katsura.
\newblock Comparing foreshock characteristics and foreshock forecasting in
  observed and simulated earthquake catalogs.
\newblock {\em Journal of Geophysical Research: Solid Earth},
  119(11):8457--8477, 2014.

\bibitem{ogata1996statistical}
Yosihiko Ogata, Tokuji Utsu, and Koichi Katsura.
\newblock Statistical discrimination of foreshocks from other earthquake
  clusters.
\newblock {\em Geophysical journal international}, 127(1):17--30, 1996.

\bibitem{parsons2014global}
Tom Parsons, Margaret Segou, and Warner Marzocchi.
\newblock The global aftershock zone.
\newblock {\em Tectonophysics}, 618:1--34, 2014.

\bibitem{srivastava2014dropout}
Nitish Srivastava, Geoffrey~E Hinton, Alex Krizhevsky, Ilya Sutskever, and
  Ruslan Salakhutdinov.
\newblock Dropout: a simple way to prevent neural networks from overfitting.
\newblock {\em Journal of Machine Learning Research}, 15(1):1929--1958, 2014.

\bibitem{wang2006predicting}
Kelin Wang, Qi-Fu Chen, Shihong Sun, and Andong Wang.
\newblock Predicting the 1975 haicheng earthquake.
\newblock {\em Bulletin of the Seismological Society of America},
  96(3):757--795, 2006.

\bibitem{wassermann2013obspy}
JM~Wassermann, L~Krischer, T~Megies, R~Barsch, and M~Beyreuther.
\newblock Obspy: a python toolbox for seismology.
\newblock In {\em AGU Fall Meeting Abstracts}, volume~1, page 2322, 2013.

\end{thebibliography}
\bibliographystyle{plain}

\end{document}